\documentclass{INTERSPEECH2023}
\usepackage{multirow}

\interspeechcameraready

\title{BA-SOT: Boundary-Aware Serialized Output Training for Multi-Talker ASR}
\name{Yuhao Liang$^1$, Fan Yu$^2$, Yangze Li$^1$, Pengcheng Guo$^1$, Shiliang Zhang$^2$, Qian Chen$^2$, Lei Xie$^{1*}$\thanks{* Corresponding author.}}
\address{$^1$Audio, Speech and Language Processing Group (ASLP@NPU), School of Computer Science, Northwestern Polytechnical University, China \\
  $^2$Speech Lab of DAMO Academy, Alibaba Group, China}
\email{liangyuhao@mail.nwpu.edu.cn, fyu@npu-aslp.org, sly.zsl@alibaba-inc.com, lxie@nwpu.edu.cn}

\begin{document}

\maketitle
\vspace{-0.4cm}
\begin{abstract}
\vspace{-0.1cm}
The recently proposed serialized output training (SOT) simplifies multi-talker automatic speech recognition (ASR) by generating speaker transcriptions separated by a special token. However, frequent speaker changes can make speaker change prediction difficult. To address this, we propose boundary-aware serialized output training (BA-SOT), which explicitly incorporates boundary knowledge into the decoder via a speaker change detection task and boundary constraint loss. We also introduce a two-stage connectionist temporal classification (CTC) strategy that incorporates token-level SOT CTC to restore temporal context information. Besides typical character error rate (CER), we introduce utterance-dependent character error rate (UD-CER) to further measure the precision of speaker change prediction. Compared to original SOT, BA-SOT reduces CER/UD-CER by 5.1\%/14.0\%, and leveraging a pre-trained ASR model for BA-SOT model initialization further reduces CER/UD-CER by 8.4\%/19.9\%.

\end{abstract}

\vspace{-0.2cm}
\section{Introduction}

Although end-to-end (E2E) ASR has been widely studied and the accuracy has reached to human-level in some general domains, multi-talker ASR in multi-party meetings scenarios is still considered a challenging problem due to the speech overlaps, noises as well as reverberation.

Recently, there has been a lot of exploration in the field of multi-party meetings scenarios~\cite{DBLP:conf/interspeech/WuLYWQ21, DBLP:journals/corr/abs-2210-15715, DBLP:conf/icassp/TaherianTW22, DBLP:journals/csl/SubramanianWWYY22,DBLP:conf/slt/YuZGLDLX22}. Progress has also been advanced with several challenges~\cite{fiscus2005rich,fiscus2006rich,fiscus2007rich,watanabe20b_chime,Yu2022M2MeT,Chen2022misp} and datasets~\cite{BarkerWVT18,mccowan2005ami,chen2020continuous,fu2021aishell,Yu2022Summary} specifically focusing on this field. 
One major problem of this scenario is the speech overlap. An intuitive solution is to use explicit speech separation followed by single talker ASR~\cite{DBLP:conf/interspeech/WuCLYTLLX20, DBLP:conf/icassp/ChenWCW0Y00021}, and permutation invariant training (PIT) method is widely applied in speaker-independent speech separation~\cite{DBLP:conf/icassp/YuKT017, DBLP:conf/icassp/WangW21}.
PIT is a method of training models that considers all possible permutations of output labels and then minimize the error by selecting the permutation that results in the best separation of individual speaker signals from mixed audio signals. 
It has also been extended and applied to multi-talker ASR~\cite{DBLP:conf/interspeech/YuCQ17,DBLP:journals/speech/QianCY18,DBLP:conf/interspeech/NeumannKBDH21} to handle permutation ambiguity, where the order of output labels varies between different hypotheses produced by the ASR model. PIT in multi-talker ASR has been advanced in various ways, including joint optimization~\cite{DBLP:journals/speech/QianCY18}, end-to-end approaches~\cite{DBLP:conf/acl/WatanabeRHSH18, DBLP:conf/icassp/ChangQ0W19}, and multi-channel methods~\cite{DBLP:conf/asru/ChangZQRW19}. However, there are still limitations to using PIT, such as the maximum number of speakers being limited by the number of output layers. Increasing the number of output layers can lead to an explosion in the complexity of the training process due to the computational complexity of O($S^3$) in PIT. Unlike PIT, conditional chain~\cite{DBLP:conf/interspeech/0003XF0020} is another popular method to build a speech separation that can effectively produce individual speaker signals from mixed audio signals with no limitation of the speaker number. By using a conditional chain, a mixed audio input with multiple speakers can be sequentially separated into individual outputs, where each output corresponds to a different speaker, with the previous output sequence serving as the conditional input. This approach can also be extended to multi-talker ASR tasks to produce the transcription of different speakers directly~\cite{DBLP:conf/nips/0003CG0F00X20,DBLP:conf/interspeech/GuoC0021}. The limitation of the conditional chain approach is computationally expensive, as it requires multiple iterations to separate each speaker signal from the mixed audio signal. The performance of the model may also be limited by the accuracy of the initial separation, as any errors introduced in the initial separation step can propagate through the subsequent steps of the conditional chain. Serialized output training (SOT)~\cite{DBLP:conf/interspeech/KandaGWMY20}, an end-to-end training method, overcomes these limitations by using a single output layer to generate the transcriptions of an unlimited number of speakers in one decoding process. 
\begin{figure*}[!ht]
	\centering
	\includegraphics[width=1.0\linewidth]{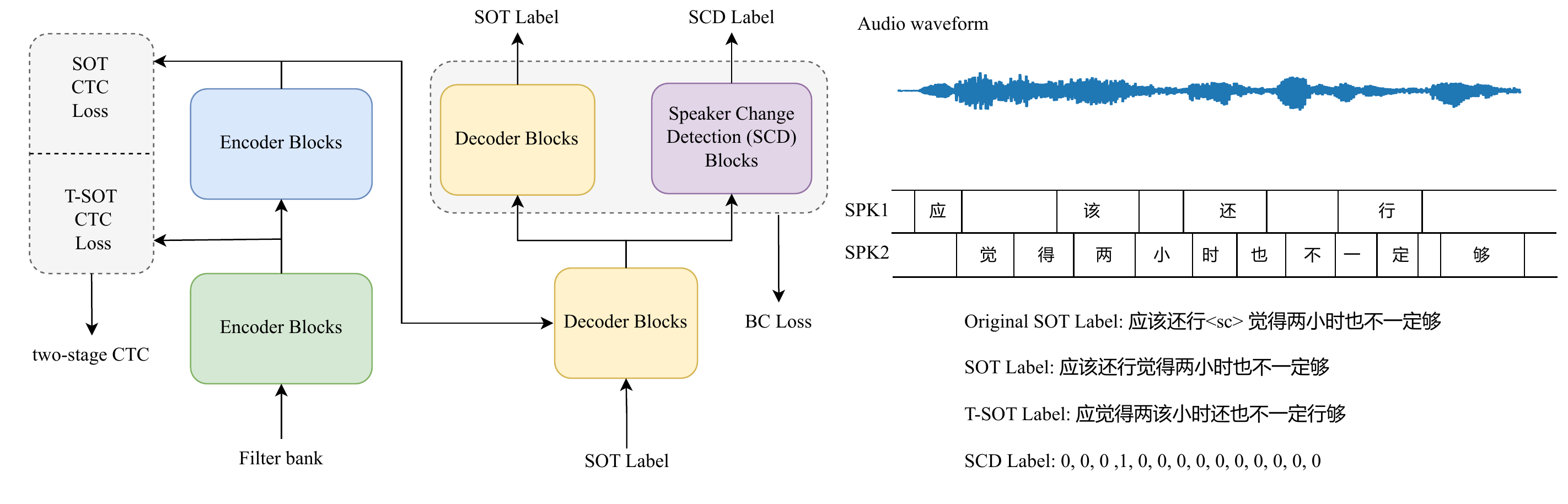}
        \vspace{-0.8cm}
	\caption{
	The architecture of our proposed method (BA-SOT) is shown on the left. And the right part of this figure shows an example of the different labels used during training and the corresponding audio waveform containing two speakers. The boundary of each token is also present and assigned to two channels according to the speaker label. 
	}
	\label{fig:model_arch}
\vspace{-0.6cm}
\end{figure*}

Despite its strengths, SOT faces two primary challenges. 
Firstly, the prediction of the special ``speaker change" token $\langle \text{sc} \rangle$ relies on the speaker's knowledge while ASR itself is a speaker-independent task. As a result, accurately predicting speaker change tokens can be difficult for typical ASR decoders. To overcome this challenge, we re-shape SOT as a multi-task learning framework with two tasks, i.e., ASR and speaker change detection (SCD). An additional SCD block is designed to capture dedicated speaker knowledge. By leveraging multi-task learning, we can learn the two tasks simultaneously and achieve a substantial improvement in the accuracy of SCD. 
The second challenge faced in the SOT is the lack of temporal monotonicity, as transcriptions are organized based on a first-in, first-out (FIFO) order. However, this goes against the temporal monotonicity mechanism of connectionist temporal classification (CTC). To solve this problem, t-SOT~\cite{DBLP:conf/interspeech/KandaWWXMWGC0Y22} arranges transcriptions in a token-level chronological order based on their emission times, but this disrupts semantic coherence and contextual information. To restore the lost temporal context information of SOT, we propose a two-stage CTC strategy that performs hierarchical encoding with the t-SOT CTC. 
Additionally, the SOT format includes audio segments that comprise multiple utterances. By providing oracle timestamps at the level of each utterance, valuable auxiliary information can be obtained for both ASR and SCD tasks. Thus, we further introduce a constrained loss that incorporates the oracle timestamp to aid both tasks.

Our proposed boundary-aware serialized output training (BA-SOT) approach preserves the capability of SOT in recognizing overlapping speech with any number of speakers while simultaneously improving the accuracy of predicting speaker changes. Experiments on real-recorded AliMeeting corpus~\cite{Yu2022M2MeT} show that BA-SOT brings a 5.1\%/14.0\% relative reduction in CER/UD-CER compared with original SOT. Furthermore, we observe that adopting a pre-trained model for BA-SOT model initiation can lead to further improvements in performance. Specifically, building the BA-SOT model on a 10000-hour-speech pre-trained ASR model~\cite{DBLP:conf/icassp/ZhangLGSYXXBCZW22} results in a relative reduction of 8.4\% and 19.9\% in CER and UD-CER, respectively.

\vspace{-0.3cm}
\section{Proposed Method}
\vspace{-0.1cm}
In this section, we present our BA-SOT based on the SOT with substantial improvements to obtain a more accurate prediction of speaker change. Specifically, Our BA-SOT includes an SCD block, a boundary constraint loss, and a two-stage CTC. Further details on each component are provided below.
\vspace{-0.3cm}
\subsection{SCD block} 
\vspace{-0.1cm}
The SOT inserts speaker change tokens between utterances to transcribe and predict the change points of different speakers. However, the success of predicting speaker change tokens depends on speaker knowledge, which is useless for the ASR task. To address this, we added a speaker change detection (SCD) block to enable multi-task learning. With this approach, two output layers are adopted to predict transcripts and speaker change separately. 
The SOT labels are modified for the ASR task to remove speaker change tokens and only transcribe speech content. The SCD task predicts sentence boundaries for each token.
This allows the ASR decoder blocks to focus on contextual information, while the SCD blocks can capture speaker knowledge individually.
\vspace{-0.1cm}
\begin{figure}[h]
	\centering
	\includegraphics[width=1.0\linewidth]{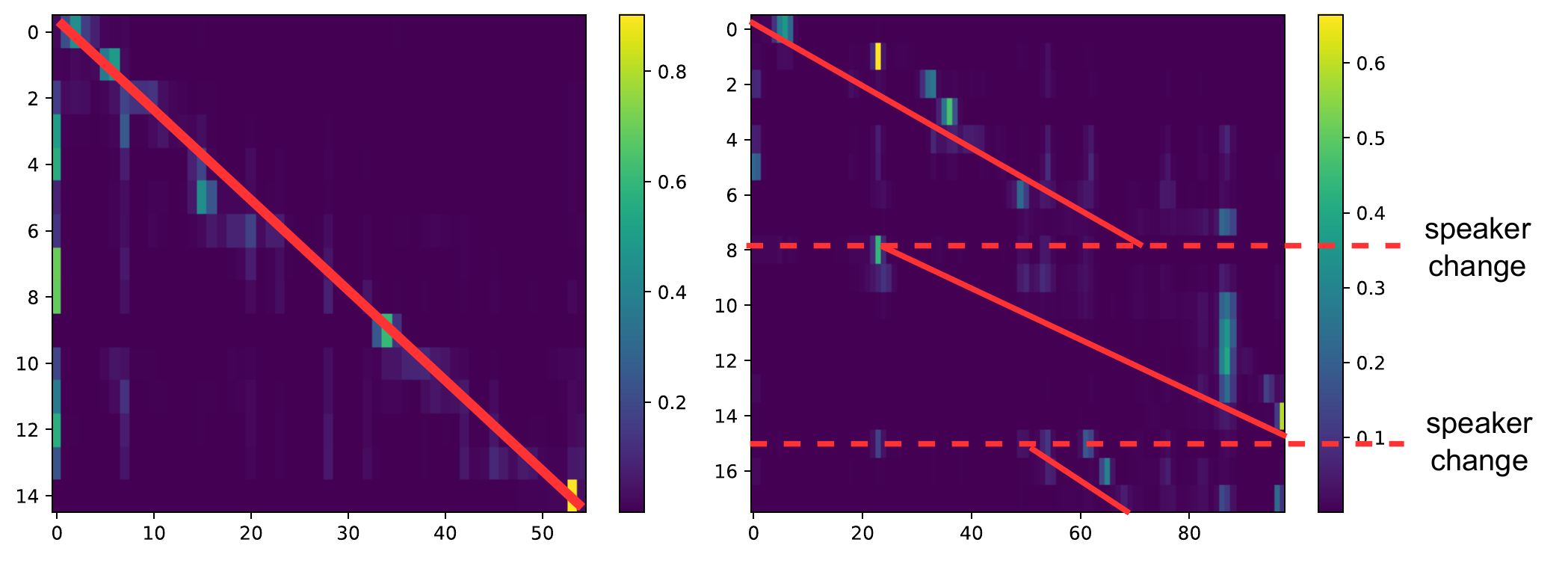}
	\caption{
	The attention map displays the expected trends of the attention score in both non-overlapping and overlapping scenarios, with the red line representing the estimated trends.}
	\label{fig:att1}
\vspace{-0.4cm}
\end{figure}
\vspace{-0.3cm}
\subsection{Boundary constraint loss}
\vspace{-0.2cm}
In order to improve both the ASR and SCD tasks, we utilize an additional loss function, called the boundary constraint (BC) loss, which is applied to the cross-attention between the encoder and decoder. For the ASR task, the BC loss leverages the known boundaries of each utterance to guide the cross-attention to focus on the correct time frames. As illustrated in Figure \ref{fig:att1} on the left, for non-overlapping speech, the attention map appears as a straight line from the upper left corner to the lower right corner. However, for overlapping speech, as shown on the right side of Figure \ref{fig:att1}, the attention line is broken into several segments, with the breaks indicating the positions of speaker changes.
Therefore, the breakpoints in the attention map provide valuable information for the SCD task. These breakpoints can be used as a substitute for speaker knowledge, and it is possible to improve the SCD task performance by encouraging the cross-attention to learn the correct boundaries using the BC loss.


Two types of BC loss are investigated as follows. For example, a SOT reference consisting of $N$ sentences and the speaker change token has been removed can be given as $\{y_{1}^{1}, ..., y_{L^1}^{1}, ..., y_{1}^{N}, ..., y_{L^N}^{N}\}$ where $y_{k}^{n}$ means the $k$-th token in $n$-th utterance and $L^n$ represent the token number of $n$-th utterance.

If oracle timestamp (OTS) is not used, the only confirmable clue is that the last token of the previous utterance usually comes after the first token of the current utterance in terms of time perspective when the speaker change token appears in SOT. Due to the lack of OTS, the timestamp from the cross attention score is adopted as a substitute. Let the predicted timestamp $t_{pre}(k, n) = \mathop{\arg\max}_{1 \leq t \leq T}(S(k, n, t))$, where $S(k, n, t)$ represents the attention score of $y_{k}^{n}$ in the time frame of $t$. Accordingly, the BC loss without OTS is formulated as:
\begin{equation}
\mathcal{L}_{\text {BC}}=
\frac{\sum\limits_{n=1}^{N}\tan\left ({\dfrac{\pi }{2}\times\max\limits_{t_{pre}(L^{n}, n)\leq t\leq T} S(1, n+1, t)} \right )}
{N}.
\end{equation}
In situations where one person dominates the conversation for an extended period and others only briefly interject, the aforementioned clue becomes less relevant. However, the use of BC loss can still provide assistance in cases where the shorter transcriptions cannot be precisely aligned, as attention tends to concentrate on either the beginning or end of the sentence. As a result, this issue has a minimal effect on the overall accuracy of the ASR output.
As the BC loss relies on the attention score's prediction of timestamps, errors can accumulate and affect performance. To address this problem, we introduce the OTS to improve the accuracy of the BC loss.
If OTS is available, the boundaries of each utterance can be determined, and attention can be guided to the correct time range. In this case, the BC loss can be defined as follows.
\begin{equation}
\mathcal{L}_{\text {BC\_OTS}}=
\frac{\sum\limits_{n=1}^{N}\sum\limits_{k=1}^{L^n}\tan\left ({\dfrac{\pi }{2}\times\max\limits_{t_{s}(n)\leq t\leq t_{e}(n)}S(k, n, t)} \right )}
{\sum\limits_{n=1}^{N}L^n},
\end{equation}
where $t_s(n)$ and $t_e(n)$ represent the start and end time of $n$-th utterance, respectively. 

\vspace{-0.3cm}
\subsection{Two-stage CTC}
\vspace{-0.2cm}
Based on the temporal monotonicity of t-SOT, we propose a two-stage CTC strategy, inspired by joint CTC/attention models used in text-to-text translation (MT) and speech-to-text translation (ST)~\cite{DBLP:journals/corr/abs-2210-05200}. The encoder is divided into two parts. The primary encoder blocks extract the acoustic feature chronologically using a t-SOT CTC loss. In this CTC loss, the separator between speakers is removed from the target label as it is redundant for speaker change prediction. The remaining encoder blocks capture the speaker's knowledge to aggregate the acoustic feature from the same speaker. They act as re-ordering layers that transfer the acoustic feature order from t-SOT to SOT format using SOT CTC loss on the output of the last layer. T-SOT is viewed as an intermediate state of SOT in this approach to alleviate the difficulty of directly modeling SOT.
\vspace{-0.3cm}
\subsection{Traning strategy}
\subsubsection{Label description}
\vspace{-0.2cm}
Our proposed method is based on SOT, which is used to serialize the transcripts of multiple overlapping sentences into a sequence and insert a speaker change token between utterances. We adopt the utterance-based first-in-first-out approach in our experiments, which rearranges the transcripts in chronological order based on their start time.

The t-SOT is also utilized to provide labels for intermediate encoding. The transcripts are sorted based on the emission times of each token and concatenated with the separator. Examples of both SOT and t-SOT are displayed on the right-hand side of Figure~\ref{fig:model_arch}.
\vspace{-0.3cm}
\subsubsection{Loss functions}
\vspace{-0.2cm}
With the BC loss working on SCD blocks, our model is jointly optimized using a multi-task objective. The standard joint CTC/attention loss is modified as follows:
\begin{equation}
\begin{aligned}
\mathcal{L}=& (1 - \lambda_{1}- \lambda_{2})\mathcal{L}_{\text{att}} + \lambda_{1}\mathcal{L}_\text{SOT\_CTC} + \lambda_{2} \mathcal{L}_{\text {t-SOT\_CTC}} \\
& + \alpha_{1}\mathcal{L}_\text{SCD} + \alpha_{2}\mathcal{L}_\text{BC/BC\_OTS},
\end{aligned}
\end{equation}
where $\mathcal{L}_{\text {att}}$ is the attention-based cross entropy, $\mathcal{L}_\text{SOT\_CTC}$ and $\mathcal{L}_\text{t-SOT\_CTC}$ is the CTC loss. In our experiment, $\lambda_{1}$ and $\lambda_{2}$ are set to 0.2, $\alpha_{1}$
is also set to 0.2 as well as $\alpha_{2}$ is set to 0.1.
\vspace{-0.3cm}
\section{Experiments}
\vspace{-0.2cm}
\subsection{Dataset}
\vspace{-0.2cm}
In this work, AliMeeting corpus, a multi-talker conversation data set in a meeting setting, is used for the evaluation of the proposed approach.
AliMeeting contains a total of 118.75 hours of speech data in total and is divided into 104.75 hours for training (Train), 4 hours for evaluation (Eval), and 10 hours for testing (Test).
The AliMeeting corpus contains far-field overlapped audios ($Ali\text{-}far$), as well as the corresponding near-field audios ($Ali\text{-}near$), which only record and transcribe the speech of a single speaker. 
The CDDMA Beamformer\cite{huang2020differential,zheng2021real} is applied to $Ali\text{-}far$ to produce $Ali\text{-}far\text{-}bf$. To evaluate the performance in a single talker scenario, $Test\_Net$, and $Test\_Meeting$ are adopted. 
$Test\_Net$ is a test set matched with WenetSpeech~\cite{DBLP:conf/icassp/ZhangLGSYXXBCZW22} corpus from the internet and $Test\_Meeting$ is another meeting dataset which is sampled from 197 real meetings in a variety of rooms. 
\vspace{-0.3cm}
\subsection{Experimental setup}
\vspace{-0.2cm}
All our experiments are conducted on ESPnet~\cite{watanabe2018espnet} toolkit. We extracted an 80-dimensional Mel-filter bank as the acoustic feature, with a frame length of 16 ms and a window shift of 8 ms.
The SOT baseline trained without model initialization comprises a 12-layer encoder and an 8-layer decoder with 4-head MHSA to ensure similar parameters compare to the SOT with SCD blocks. The dimensions of the MHSA and FFN layers are set to 256 and 2048, respectively. 
We utilized 2 transformer layers with the same configuration as the decoder blocks for the SCD blocks. As for the two-stage CTC, we employed 9 encoder layers to extract acoustic features in t-SOT order and followed it with 3 encoder layers to accomplish re-ordering.
The absence of speaker changes in the $Ali\text{-}near$ dataset would exacerbate the issue of class imbalance and adversely affect the accuracy of our model. Therefore, we perform fine-tuning of the SCD blocks exclusively on the far-field dataset after completing the model training.
\vspace{-0.3cm}
\subsection{Evaluation metric}
\vspace{-0.1cm}
In a multi-talker ASR system, the number of hypothesis generated may differ from the number of references, which can indicate inaccurate speaker change prediction. To address this issue, we introduce utterance-dependent CER (UD-CER), which involves dividing the hypotheses by speaker change token into multiple utterances and rearranging them to obtain the lowest CER. The calculation of UD-CER is similar to that of the Optimal Reference Combination (ORC) WER and it can be seen as a variant of ORC WER designed for N channels, where N represents the number of utterances. There are some distinctions between UD-CER and the WER definitions used in the original SOT. While both metrics permute the utterances, the WER in SOT concatenates all utterances and then computes CER once. It is worth noting that UD-CER is particularly sensitive to errors in predicting speaker changes. Besides UD-CER, we also utilize the character error rate (CER) metric following the protocol established in the M2MeT challenge.

\begin{table*}[h]
\caption{CER/UD-CER of our proposed method on several test sets. SOT baseline with model initialization random initializes the output layer due to the dictionary difference caused by the special token $\langle \text{sc} \rangle$. All of the models are decoded without language model and the CTC weight during decoding is set to 0.6. The beam size is set to 20.}
\vspace{-0.2cm}
\setlength{\tabcolsep}{1.4mm}{
\renewcommand{\arraystretch}{1.16}
\begin{tabular}{l|lccccc}
\toprule
                                    & Models                        & Param (M) & Eval-Ali-far-bf & Test-Ali-far-bf & Test\_Meeting    & Test\_Net        \\ \midrule
\multirow{6}{*}{w/o model init}  & SOT (baseline)                & 48.53     & 30.8/34.8       & 29.7/35.6       & \textbackslash{} & \textbackslash{} \\ \cline{2-7} 
                                    &\quad + two-stage CTC & 49.47     & 29.9/33.8       & 29.5/35.2       & \textbackslash{} & \textbackslash{} \\ \cline{2-7} 
                                    &\qquad + BC loss with OTS            & 49.47     & 29.0/32.1       & 28.9/32.0       & \textbackslash{} & \textbackslash{} \\ \cline{2-7} 
                                    &\qquad + SCD block                    & 49.47     & 29.9/35.7       & 29.6/35.3      & \textbackslash{} & \textbackslash{} \\ \cline{2-7} 
                                    &\quad\qquad + BC loss                     & 49.47     & 29.0/33.0       & 29.2/33.3       & \textbackslash{} & \textbackslash{} \\ \cline{2-7} 
                                    &\quad\qquad + BC loss with OTS            & 49.47     & \textbf{28.1}/\textbf{30.2}       & \textbf{28.2}/\textbf{30.6}       & \textbackslash{} & \textbackslash{} \\ \midrule
\multirow{5}{*}{w/ model init} & pre-trained model                    & 116.9     & 31.2/           & 30.6/           & 15.8             & \textbf{8.7}              \\ \cline{2-7} 
                                    & SOT (baseline)                & 116.9     & 23.1/28.5       & 22.5/28.6       & 9.3              & 10.8             \\ \cline{2-7} 
                                    &\quad + two-stage CTC & 119.7     & 22.9/28.2       & 22.2/28.2       & 9.1              & 10.7             \\ \cline{2-7} 
                                    &\qquad + SCD block \& BC loss              & 128.1     & 22.4/26.5       & 21.8/26.0       & 9.1              & 10.9             \\ \cline{2-7} 
                                    &\qquad + SCD block \& BC loss with OTS     & 128.1     & \textbf{21.3}/\textbf{23.7}       & \textbf{20.6}/\textbf{22.9}       & \textbf{8.8}              & 10.5             \\ \bottomrule
\end{tabular}}
\label{tab:result}
\vspace{-0.5cm}
\end{table*}
\vspace{-0.3cm}
\subsection{Results without model initialization}
\vspace{-0.1cm}
As shown in Table~\ref{tab:result}, we compare the effect of our proposed methods with those of the SOT baseline model, which is trained from scratch on $Train\text{-}Ali\text{-}far\text{-}bf$.
Two-stage CTC decreases the CER and UD-CER from 29.7\%/35.6\% to 29.5\%/35.2\% on the test set and BC loss with OTS can further reduce the CER and UD-CER to 28.9\%/32.0\%. 
The SCD block increases the UD-CER when it works without BC loss. With the help of BC loss, the SCD block can achieve better prediction performance and further reduce the CER compared to using BC loss alone. 
Our proposed BA-SOT, leads to 5.1\% related CER reduction (29.7\% $\to$ 28.2\%) and 14.0\% related UD-CER reduction (35.6\% $\to$ 30.6\%) on the test set.

\vspace{-0.3cm}
\subsection{Results with model initialization}
\vspace{-0.1cm}
The BC loss  without an oracle timestamp tends to confuse the model by optimizing in the wrong direction because of the random initialization of the model parameters. Therefore, a pre-trained model$\footnote{https://github.com/espnet/espnet/tree/master/egs2/wenetspeech/asr1}$ is employed to provide a desirable initialization. 
$Train\text{-}Ali\text{-}far\text{-}bf$ and $Train\text{-}Ali\text{-}near$ are used to investigate our approaches by training with the pre-trained model initialization, and the results are shown in Table~\ref{tab:result}. 
The SOT model can significantly improve the performance of the datasets of meeting scenarios. Specifically, the SOT baseline fine-tuned on the pre-trained ASR model substantially decreases the CER from 31.2\%, 30.6\%, and 15.8\% to 23.1\%, 22.5\% and 9.3\% on $Eval\text{-}Ali\text{-}far\text{-}bf$, $Test\text{-}Ali\text{-}far\text{-}bf$ and $Test\_Meeting$, respectively.
However, SOT increases the CER on $Test\_Net$, from 8.7\% to 10.8\% which may be caused by the domain mismatch between AliMeeting and WenetSpeech corpus. On top of this, our proposed methods can further decrease the CER and UD-CER that achieves 8.4\%/19.9\% related CER/UD-CER reduction (22.5\%/28.6\% $\to$ 20.6\%/22.9\%) on $Test\text{-}Ali\text{-}far\text{-}bf$. 
Moreover, our proposed method outperformed fine-tuned SOT for the single-talker ASR condition on $Test\_Net$ (10.5\% vs. 10.8\%).



\vspace{-0.3cm}
\subsection{Analysis of the attention map}
\begin{figure}[h]
	\centering
	\includegraphics[width=1.0\linewidth]{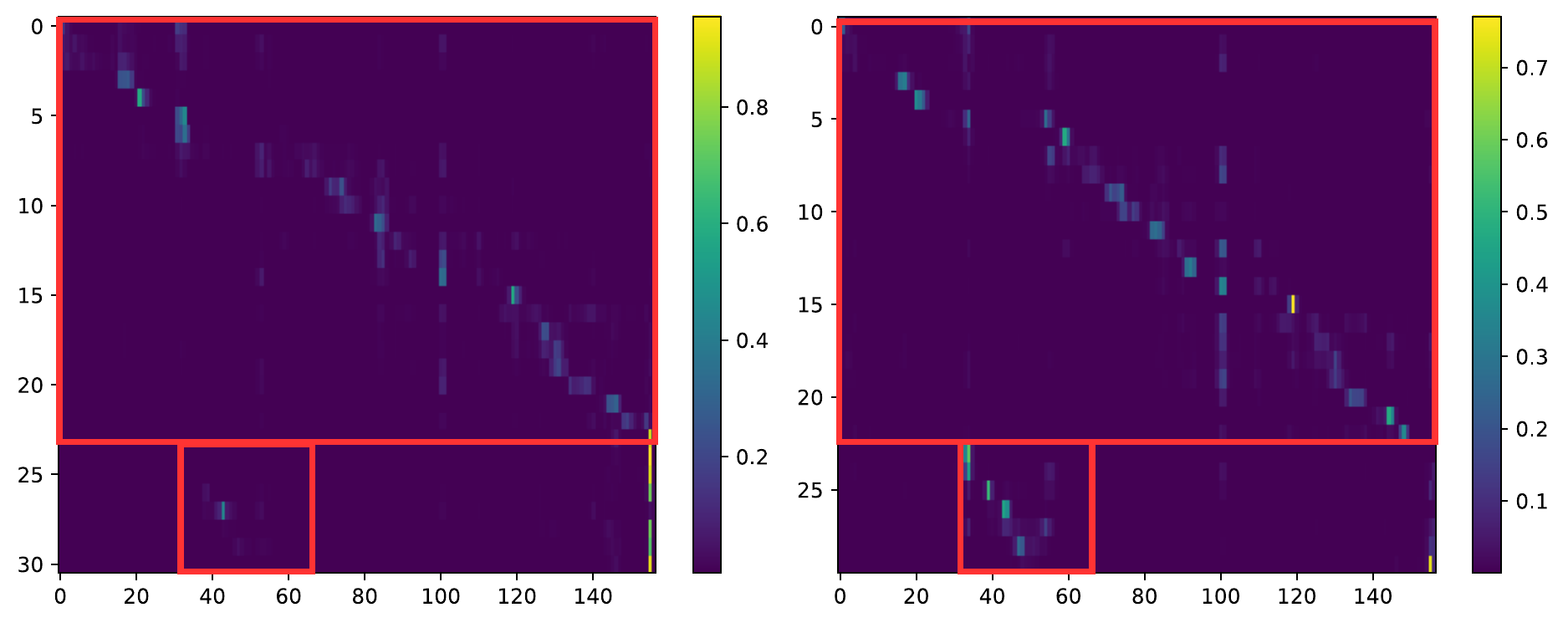}
\vspace{-0.5cm}
	\caption{
	The figure displays the attention map for the audio segment R8002\_M8003\_MS803-0015074-0015579 in the AliMeeting corpus. The red square box denotes the temporal arrangement of each utterance.
	}
	\label{fig:att2}
\vspace{-0.3cm}
\end{figure}

\vspace{-0.1cm}
Figure~\ref{fig:att2} presents the attention maps of the SOT model. The left part of the figure shows that the attention is confused at the actual time boundary of the corresponding utterance. In contrast, the right part displays the attention map of the BA-SOT, which is more focused and clear. This improvement is achieved through the use of the BC loss, which helps the attention not only accelerate the convergence speed of the model but also enhance the attention accuracy of the label boundary
\vspace{-0.2cm}
\vspace{-0.4cm}
\subsection{Analysis of speaker change detection accuracy}
\begin{figure}[h]
	\centering
	\includegraphics[width=1.0\linewidth]{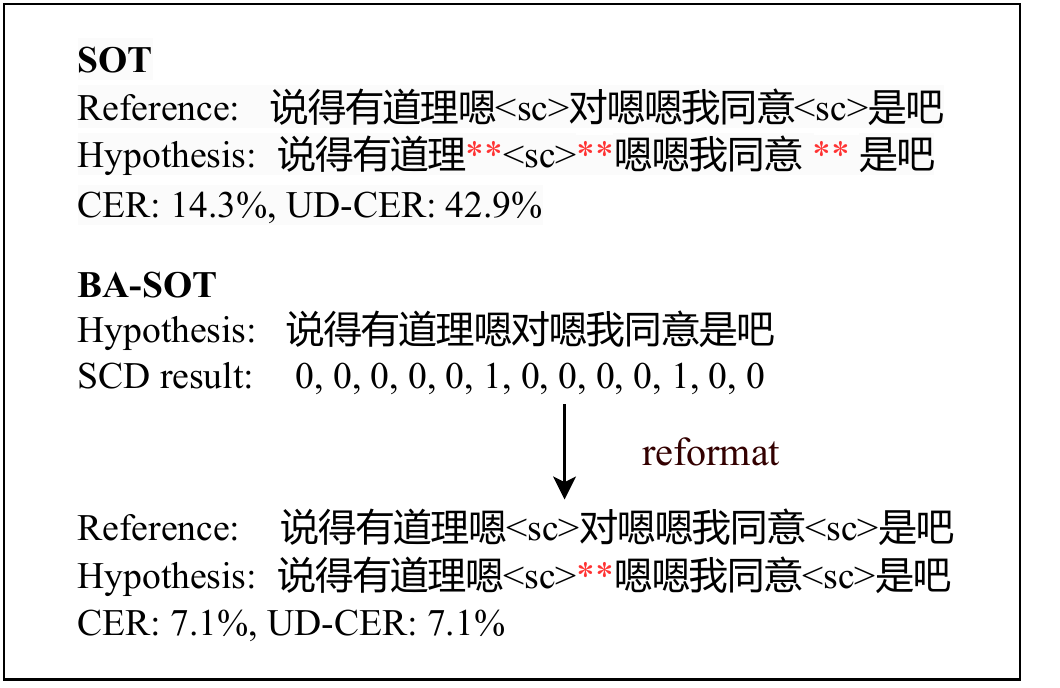}
\vspace{-0.4cm}
	\caption{
	A example of decoding result. The red *** indicates a delete or insert error. The output of BA-SOT is reformatted to SOT format and then the CER and UD-CER are calculated.
	}
	\label{fig:hyp}
\vspace{-0.2cm}
\end{figure}

\vspace{-0.1cm}
A decoding result is shown in Figure ~\ref{fig:hyp}. The hypothesis produced by SOT misses one speaker change token, resulting in the soaring of UD-CER and BA-SOT can achieve a better prediction on the speaker change. 
In this example, the BA-SOT results indicate that when predicting speaker change positions with higher precision, UD-CER can equal CER. The trends of CER and UD-CER in Table~\ref{tab:result} and Figure~\ref{fig:hyp} are similar, with UD-CER showing more significant improvement than CER. This suggests that our proposed method can more accurately predict speaker change positions.
\vspace{-0.2cm}
\vspace{-0.3cm}
\section{Conclusions}
\vspace{-0.1cm}
This paper proposes a boundary-aware serialized output training approach to improve the accuracy of speaker change prediction in SOT. Incorrectly predicting speaker change can have serious consequences on the accuracy of multi-talker ASR. To solve with this problem, we employ SCD blocks for multi-task learning and apply the BC loss to the cross attention to enforce correct boundary and speaker change detection in overlapped speech. We then adopt the two-stage CTC, which uses an additional t-SOT CTC loss. Although individual use of SCD blocks may have negative effects, the BC loss further reduces CER. Our proposed method achieves a 5.1\%/14.0\% related CER/UD-CER reduction on the test set of AliMeeting corpus compared to the standard SOT. With pre-trained ASR model initialization, our approaches achieve an 8.4\%/19.9\% related CER/UD-CER reduction on the Test-Ali-far-bf dataset while maintaining a comparable CER of 10.5\% in single talker conditions. 

\bibliographystyle{IEEEtran}
\bibliography{mybib}

\begin{thebibliography}{10}
\providecommand{\url}[1]{#1}
\csname url@samestyle\endcsname
\providecommand{\newblock}{\relax}
\providecommand{\bibinfo}[2]{#2}
\providecommand{\BIBentrySTDinterwordspacing}{\spaceskip=0pt\relax}
\providecommand{\BIBentryALTinterwordstretchfactor}{4}
\providecommand{\BIBentryALTinterwordspacing}{\spaceskip=\fontdimen2\font plus
\BIBentryALTinterwordstretchfactor\fontdimen3\font minus
  \fontdimen4\font\relax}
\providecommand{\BIBforeignlanguage}[2]{{%
\expandafter\ifx\csname l@#1\endcsname\relax
\typeout{** WARNING: IEEEtran.bst: No hyphenation pattern has been}%
\typeout{** loaded for the language `#1'. Using the pattern for}%
\typeout{** the default language instead.}%
\else
\language=\csname l@#1\endcsname
\fi
#2}}
\providecommand{\BIBdecl}{\relax}
\BIBdecl

\bibitem{DBLP:conf/interspeech/WuLYWQ21}
Y.~Wu, C.~Li, S.~Yang, Z.~Wu, and Y.~Qian, ``Audio-visual multi-talker speech
  recognition in a cocktail party,'' in \emph{Proc. Interspeech 2021}.\hskip
  1em plus 0.5em minus 0.4em\relax {ISCA}, 2021, pp. 3021--3025.

\bibitem{DBLP:journals/corr/abs-2210-15715}
M.~Yang, N.~Kanda, X.~Wang, J.~Wu, S.~Sivasankaran, Z.~Chen, J.~Li, and
  T.~Yoshioka, ``Simulating realistic speech overlaps improves multi-talker
  {ASR},'' \emph{CoRR}, vol. abs/2210.15715, 2022.

\bibitem{DBLP:conf/icassp/TaherianTW22}
H.~Taherian, K.~Tan, and D.~Wang, ``Location-based training for multi-channel
  talker-independent speaker separation,'' in \emph{Proc. ICASSP}.\hskip 1em
  plus 0.5em minus 0.4em\relax {IEEE}, 2022, pp. 696--700.

\bibitem{DBLP:journals/csl/SubramanianWWYY22}
A.~S. Subramanian, C.~Weng, S.~Watanabe, M.~Yu, and D.~Yu, ``Deep learning
  based multi-source localization with source splitting and its effectiveness
  in multi-talker speech recognition,'' \emph{Comput. Speech Lang.}, vol.~75,
  p. 101360, 2022.

\bibitem{DBLP:conf/slt/YuZGLDLX22}
F.~Yu, S.~Zhang, P.~Guo, Y.~Liang, Z.~Du, Y.~Lin, and L.~Xie,
  ``{MFCCA}:multi-frame cross-channel attention for multi-speaker {ASR} in
  multi-party meeting scenario,'' in \emph{Proc. SLT}.\hskip 1em plus 0.5em
  minus 0.4em\relax {IEEE}, 2022, pp. 144--151.

\bibitem{fiscus2005rich}
J.~G. Fiscus, N.~Radde, J.~S. Garofolo, A.~Le, J.~Ajot, and C.~Laprun, ``The
  rich transcription 2005 spring meeting recognition evaluation,'' in
  \emph{Proc. MLMI}.\hskip 1em plus 0.5em minus 0.4em\relax Springer, 2005, pp.
  369--389.

\bibitem{fiscus2006rich}
J.~G. Fiscus, J.~Ajot, M.~Michel, and J.~S. Garofolo, ``The rich transcription
  2006 spring meeting recognition evaluation,'' in \emph{Proc. MLMI}.\hskip 1em
  plus 0.5em minus 0.4em\relax Springer, 2006, pp. 309--322.

\bibitem{fiscus2007rich}
J.~G. Fiscus, J.~Ajot, and J.~S. Garofolo, ``The rich transcription 2007
  meeting recognition evaluation,'' in \emph{Proc. MTPH}.\hskip 1em plus 0.5em
  minus 0.4em\relax Springer, 2007, pp. 373--389.

\bibitem{watanabe20b_chime}
S.~Watanabe, M.~Mandel, J.~Barker \emph{et~al.}, ``{CHiME-6 Challenge: Tackling
  Multispeaker Speech Recognition for Unsegmented Recordings},'' in \emph{Proc.
  CHiME}, 2020, pp. 1--7.

\bibitem{Yu2022M2MeT}
F.~Yu, S.~Zhang, Y.~Fu, L.~Xie, S.~Zheng, Z.~Du \emph{et~al.}, ``M2{M}e{T}: The
  {ICASSP} 2022 multi-channel multi-party meeting transcription challenge,'' in
  \emph{Proc. ICASSP}.\hskip 1em plus 0.5em minus 0.4em\relax IEEE, 2022.

\bibitem{Chen2022misp}
H.~Chen, H.~Zhou, J.~Du, C.-H. Lee, J.~Chen, S.~Watanabe, S.~M. Siniscalchi,
  O.~Scharenborg, D.-Y. Liu, B.-C. Yin, J.~Pan, J.-Q. Gao, and C.~Liu, ``The
  first multimodal information based speech processing ({MISP}) challenge:
  Data, tasks, baselines and results,'' in \emph{Proc. ICASSP}.\hskip 1em plus
  0.5em minus 0.4em\relax IEEE, 2022, pp. 9266--9270.

\bibitem{BarkerWVT18}
J.~Barker, S.~Watanabe, E.~Vincent, and J.~Trmal, ``The fifth '{CHiME}' speech
  separation and recognition challenge: Dataset, task and baselines,'' in
  \emph{Proc. INTERSPEECH}.\hskip 1em plus 0.5em minus 0.4em\relax ISCA, 2018,
  pp. 1561--1565.

\bibitem{mccowan2005ami}
I.~McCowan, J.~Carletta, W.~Kraaij, S.~Ashby, S.~Bourban, M.~Flynn
  \emph{et~al.}, ``The {AMI} meeting corpus,'' in \emph{Proc. ICMT}, vol.~88,
  2005, p. 100.

\bibitem{chen2020continuous}
Z.~Chen, T.~Yoshioka, L.~Lu, T.~Zhou, Z.~Meng, Y.~Luo, J.~Wu, X.~Xiao, and
  J.~Li, ``Continuous speech separation: {D}ataset and analysis,'' in
  \emph{Proc. ICASSP}.\hskip 1em plus 0.5em minus 0.4em\relax IEEE, 2020, pp.
  7284--7288.

\bibitem{fu2021aishell}
Y.~Fu, L.~Cheng, S.~Lv, Y.~Jv, Y.~Kong, Z.~Chen, Y.~Hu \emph{et~al.},
  ``{AISHELL}-4: {A}n open source dataset for speech enhancement, separation,
  recognition and speaker diarization in conference scenario,'' in \emph{Proc.
  INTERSPEECH}.\hskip 1em plus 0.5em minus 0.4em\relax ISCA, 2021, pp.
  3665--3669.

\bibitem{Yu2022Summary}
F.~Yu, S.~Zhang, P.~Guo, Y.~Fu, Z.~Du, S.~Zheng, L.~Xie \emph{et~al.},
  ``Summary on the {ICASSP} 2022 multi-channel multi-party meeting
  transcription grand challenge,'' in \emph{Proc. ICASSP}.\hskip 1em plus 0.5em
  minus 0.4em\relax IEEE, 2022.

\bibitem{DBLP:conf/interspeech/WuCLYTLLX20}
J.~Wu, Z.~Chen, J.~Li, T.~Yoshioka, Z.~Tan, E.~Lin, Y.~Luo, and L.~Xie, ``An
  end-to-end architecture of online multi-channel speech separation,'' in
  \emph{Proc. Interspeech}.\hskip 1em plus 0.5em minus 0.4em\relax {ISCA},
  2020, pp. 81--85.

\bibitem{DBLP:conf/icassp/ChenWCW0Y00021}
S.~Chen, Y.~Wu, Z.~Chen, J.~Wu, J.~Li, T.~Yoshioka, C.~Wang, S.~Liu, and
  M.~Zhou, ``Continuous speech separation with conformer,'' in \emph{Proc.
  ICASSP}.\hskip 1em plus 0.5em minus 0.4em\relax {IEEE}, 2021, pp. 5749--5753.

\bibitem{DBLP:conf/icassp/YuKT017}
D.~Yu, M.~Kolb{\ae}k, Z.~Tan, and J.~Jensen, ``Permutation invariant training
  of deep models for speaker-independent multi-talker speech separation,'' in
  \emph{Proc. ICASSP}.\hskip 1em plus 0.5em minus 0.4em\relax {IEEE}, 2017, pp.
  241--245.

\bibitem{DBLP:conf/icassp/WangW21}
Z.~Wang and D.~Wang, ``Count and separate: Incorporating speaker counting for
  continuous speaker separation,'' in \emph{Proc. ICASSP}.\hskip 1em plus 0.5em
  minus 0.4em\relax {IEEE}, 2021, pp. 11--15.

\bibitem{DBLP:conf/interspeech/YuCQ17}
D.~Yu, X.~Chang, and Y.~Qian, ``Recognizing multi-talker speech with
  permutation invariant training,'' in \emph{Proc. Interspeech}.\hskip 1em plus
  0.5em minus 0.4em\relax {ISCA}, 2017, pp. 2456--2460.

\bibitem{DBLP:journals/speech/QianCY18}
Y.~Qian, X.~Chang, and D.~Yu, ``Single-channel multi-talker speech recognition
  with permutation invariant training,'' \emph{Speech Commun.}, pp. 1--11,
  2018.

\bibitem{DBLP:conf/interspeech/NeumannKBDH21}
T.~von Neumann, K.~Kinoshita, C.~B{\"{o}}ddeker, M.~Delcroix, and
  R.~Haeb{-}Umbach, ``Graph-pit: Generalized permutation invariant training for
  continuous separation of arbitrary numbers of speakers,'' in \emph{Proc.
  Interspeech}.\hskip 1em plus 0.5em minus 0.4em\relax {ISCA}, 2021, pp.
  3490--3494.

\bibitem{DBLP:conf/acl/WatanabeRHSH18}
H.~Seki, T.~Hori, S.~Watanabe, J.~L. Roux, and J.~R. Hershey, ``A purely
  end-to-end system for multi-speaker speech recognition,'' in \emph{Proc.
  ACL}.\hskip 1em plus 0.5em minus 0.4em\relax ACL, 2018, pp. 2620--2630.

\bibitem{DBLP:conf/icassp/ChangQ0W19}
X.~Chang, Y.~Qian, K.~Yu, and S.~Watanabe, ``End-to-end monaural multi-speaker
  {ASR} system without pretraining,'' in \emph{proc. ICASSP}.\hskip 1em plus
  0.5em minus 0.4em\relax {IEEE}, 2019, pp. 6256--6260.

\bibitem{DBLP:conf/asru/ChangZQRW19}
X.~Chang, W.~Zhang, Y.~Qian, J.~L. Roux, and S.~Watanabe, ``Mimo-speech:
  End-to-end multi-channel multi-speaker speech recognition,'' in \emph{Proc.
  ASRU}.\hskip 1em plus 0.5em minus 0.4em\relax {IEEE}, 2019, pp. 237--244.

\bibitem{DBLP:conf/interspeech/0003XF0020}
J.~Shi, J.~Xu, Y.~Fujita, S.~Watanabe, and B.~Xu, ``Speaker-conditional chain
  model for speech separation and extraction,'' in \emph{Proc.
  Interspeech}.\hskip 1em plus 0.5em minus 0.4em\relax {ISCA}, 2020, pp.
  2707--2711.

\bibitem{DBLP:conf/nips/0003CG0F00X20}
J.~Shi, X.~Chang, P.~Guo, S.~Watanabe, Y.~Fujita, J.~Xu, B.~Xu, and L.~Xie,
  ``Sequence to multi-sequence learning via conditional chain mapping for
  mixture signals,'' in \emph{Proc. NIPS}, 2020.

\bibitem{DBLP:conf/interspeech/GuoC0021}
P.~Guo, X.~Chang, S.~Watanabe, and L.~Xie, ``Multi-speaker {ASR} combining
  non-autoregressive conformer {CTC} and conditional speaker chain,'' in
  \emph{Proc. Interspeech}.\hskip 1em plus 0.5em minus 0.4em\relax {ISCA},
  2021, pp. 3720--3724.

\bibitem{DBLP:conf/interspeech/KandaGWMY20}
N.~Kanda, Y.~Gaur, X.~Wang, Z.~Meng, and T.~Yoshioka, ``Serialized output
  training for end-to-end overlapped speech recognition,'' in \emph{Proc.
  Interspeech}.\hskip 1em plus 0.5em minus 0.4em\relax {ISCA}, 2020, pp.
  2797--2801.

\bibitem{DBLP:conf/interspeech/KandaWWXMWGC0Y22}
N.~Kanda, J.~Wu, Y.~Wu, X.~Xiao, Z.~Meng, X.~Wang, Y.~Gaur, Z.~Chen, J.~Li, and
  T.~Yoshioka, ``Streaming multi-talker {ASR} with token-level serialized
  output training,'' in \emph{Proc. Interspeech}.\hskip 1em plus 0.5em minus
  0.4em\relax {ISCA}, 2022, pp. 3774--3778.

\bibitem{DBLP:conf/icassp/ZhangLGSYXXBCZW22}
B.~Zhang, H.~Lv, P.~Guo, Q.~Shao, C.~Yang, L.~Xie, X.~Xu, H.~Bu, X.~Chen,
  C.~Zeng, D.~Wu, and Z.~Peng, ``{WENETSPEECH:} {A} 10000+ hours multi-domain
  mandarin corpus for speech recognition,'' in \emph{Proc. ICASSP}.\hskip 1em
  plus 0.5em minus 0.4em\relax {IEEE}, 2022, pp. 6182--6186.

\bibitem{DBLP:journals/corr/abs-2210-05200}
B.~Yan, S.~Dalmia, Y.~Higuchi, G.~Neubig, F.~Metze, A.~W. Black, and
  S.~Watanabe, ``{CTC} alignments improve autoregressive translation,'' in
  \emph{arXiv preprint arXiv:2210.05200}, 2022.

\bibitem{huang2020differential}
W.~Huang and J.~Feng, ``Differential beamforming for uniform circular array
  with directional microphones.'' in \emph{Proc. INTERSPEECH}.\hskip 1em plus
  0.5em minus 0.4em\relax ISCA, 2020, pp. 71--75.

\bibitem{zheng2021real}
S.~Zheng, W.~Huang, X.~Wang, H.~Suo, J.~Feng, and Z.~Yan, ``A real-time speaker
  diarization system based on spatial spectrum,'' in \emph{Proc. ICASSP}.\hskip
  1em plus 0.5em minus 0.4em\relax IEEE, 2021, pp. 7208--7212.

\bibitem{watanabe2018espnet}
S.~Watanabe, T.~Hori, S.~Karita, T.~Hayashi, J.~Nishitoba, Y.~Unno, N.~{Enrique
  Yalta Soplin}, J.~Heymann, M.~Wiesner, N.~Chen, A.~Renduchintala, and
  T.~Ochiai, ``{ESPnet}: End-to-end speech processing toolkit,'' in \emph{Proc.
  Interspeech}, 2018, pp. 2207--2211.

\end{thebibliography}

\end{document}